\documentclass[a4paper,10pt]{article}
\usepackage{graphicx}
\usepackage{afterpage}

\title{Conjecture on a classification criterion for holonomies for the $U(1)$ group}
\author{{Mendes}, Daniel de Cerqueira Lima\\
danim@cetuc.puc-rio.br}

\begin{document}

\maketitle

\begin{abstract}
Our main goal is to find a criterion to classify if a given group have a structure similar to a U(1) Lie group.
\end{abstract}

\bigskip

\section{The Infintesimal Logarithmic Spiral}

\paragraph{}
In the course of seeking such a criterion, lets start with the analisys of another problem. Lets start with a problem of reversibility. The Bernoulli application 

\bigskip

$$A_{i+1} = {A_{i} \over 2} \parallel 1 \parallel \eqno(1.1)$$


where $\parallel K \parallel$ stands for $mod(K)$.

\bigskip

This is an example of an irreversible transformation. In first place, the inverse transformation have an atractor in 0. Moreover, information is always lost in the module process. More generally, a transformation as

\bigskip

$$A_{i+\epsilon} = f(A_{i}) \parallel 2K\pi \parallel \eqno(1.2)$$

\bigskip

is readily indentified as an irreversible transformation. Comparing with the function $f$, our transformation may be visualised as taking the domain line and wrapping it around a circle with radius $K$. This procedure makes the transformation irreversible, even disregarding possible atractors of the function $f$ when submitted to the transformation. This happens because many points of the image of $f$ are projected over the same point in the circunference. In this case, a possible tentative of eliminating the irreversibility would be to transform our circle in an spiral wich have an infinitesimal change in its curvature ratio. The most obvious choice is the infinitesimal logarithmic spiral. So we could transform our circle in a spiral using the following diffeomorfism, in polar coordinates.

\bigskip

$$R(\theta) = K \Longrightarrow R(\theta) = e^{\epsilon\theta} \eqno(1.3)$$

\bigskip

The neighborhood of a point in the circle would be transferred to the neighborhhod of a point of the spiral in a bijective way, but analitically, our transformation would be no more \textit{a priori} irreversible.

\bigskip

\begin{figure}[h]
\centering
\includegraphics[scale=0.750]{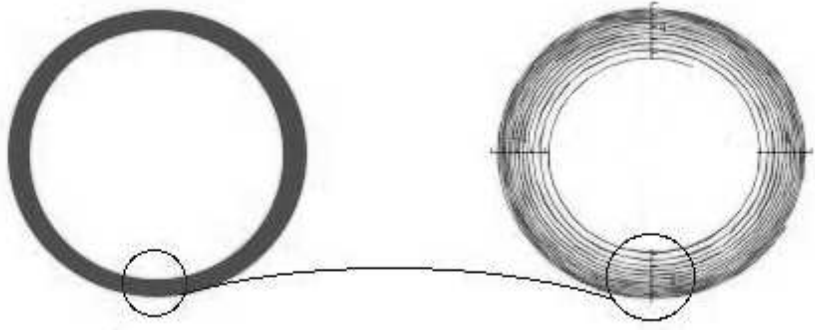}
\caption{Diffeomorfism between a circle and a logarithmic spiral}
\end{figure}

Speaking objectively, we want to study if makes sense to regard the application of a logarithmic spiral as a criterion for classifying groups of coordinate transformations as being similar to a $U(1)$ group or not.

\paragraph{}
An element of the $U(1)$ group, the group of rotations in the internal space with on real parameter, may be exponentiated as $g = e^{i\theta}$. 
The basic criterions to form a $U(1)$ group are:

$$g_{i}g_{j} \in U(1) - ex: e^{i\theta_{i}}e^{i\theta_{j}} = e^{i(\theta_{i}+\theta_{j})} = e^{i\theta_{k}} \in U(1) \eqno(1.4)$$

$$gg^{\dagger} = g^{\dagger}g = 1 - ex: e^{i\theta}e^{-i\theta} = e^{-i\theta}e^{i\theta} = 1\eqno(1.5)$$

as well as to obey to the Jacobi indentity

$$[[g_{i},g_{j}], g_{k}] + [[g_{k},g_{i}], g_{j}] + [[g_{j},g_{k}], g_{i}] = 0 \eqno(1.6)$$

where $[ x, y ]$ stands for an anticommutative bilinear in $x$ and $y$.

If we multiply our element $g$ by $e^{\epsilon\theta}$ we would have an element of the form

$$e^{\epsilon\theta}e^{i\theta} = e^{(\epsilon + i)\theta}\eqno(1.7)$$

in general, our new element is out of the $U(1)$ group, because

$$gg^{\dagger} = e^{(\epsilon + i)\theta}e^{(\epsilon - i)\theta} = e^{2\epsilon\theta} \neq 1 \longrightarrow g \notin U(1) \eqno(1.8)$$

\paragraph{}
In a specific case, however, our element stays in the $U(1)$ group. The Lie groups are continous groups, specificaly continous tranformations groups. The discrete transformation are obtained from the continous transformation. In the case where $theta$ is infinitesimal, our group element stays in the $U(1)$ group, once we are using a first order theory.

$$gg^{\dagger} = lim_{\theta\to0}e^{2\epsilon\theta} = e^{0} = 1 \longrightarrow g \in U(1) \eqno(1.9) $$

\paragraph{}
Here we cand find clearly a symmetry break between the major parts, the discrete transformations and the minor parts, the inifinitesimal transformations from wich the former are composed. Due to the definition of a continous group, we can regard the discrete transformations as composed by the inifinitesimal ones. Obviously the discrete elements can overlap, etc. So we have a simetry break between to scales of transformations wich can be said as following. 

\paragraph{}
The application of a inifinitesimal logarithmic spiral to an element of the group, defines if we are dealing with an inifinitesimal element or a discrete one. In the case of an inifinitesimal element, it stays in the group, otherwise if in the case of a finite one it gets away of the group.

\paragraph{}
The question is - Can the application of the infinitesimal logarithmic spiral transformation to a group of coordinate transformations define if is possible to make an holonomy of the group with the $U(1)$ group? 

\paragraph{}
Another question is - In a coordinate transformation group, what stands for the element in what the criterion will be applied? Our first choice is to test the Jacobian. More specifically, the determinant of the jacobian of the coordinate transformation. In the case of a finite element it would take away from the group, otherwise, in the case of an infinitesimal element it would stay in the group. If the group don't follow this rule it could not be considered an holonomy between the group and the $U(1)$ group. In specific cases other methods may be used to apply this criterion directly on the coordinate transformation group.

\paragraph{}
Finally, is there something that deserves analisys in the symmetry breaking between the finite and infinitesimal elements of the $U(1)$ and similar (in our sense) groups when \textit{operated} by the infinitesimal logarithmic spiral?




\newpage\end{document}